# An update to 'Allocating time on scientific platforms in outer space' using five cycles of JWST General Observer programs

Christopher Williams
ESSCA School of Management
55 Quai Alphonse le Gallo
92513 Boulogne-Billancourt
France

Email: christopher.williams@essca.fr

**ABSTRACT**

'Allocating time on scientific platforms in outer space' (Williams, 2025 in *Research Policy*) explored time allocated to astronomers to use the James Webb Space Telescope (JWST). It was based on data from the first three Cycles of General Observer (GO) programs (2022, 2023 and 2024) and used multi-variable regression modelling. Analysis provided support for progress towards reducing bias in time allocation decisions in terms of Principal Investigator (PI) gender, institution and country. It also provided strong evidence that higher time allocation is linked to larger teams and data non-exclusivity. However, a question was raised concerning a possible swing towards long-standing PI gender bias as the Cycles progressed, i.e., that equality might only be temporary. In addition, the possibility of a Trump-era effect on time allocation bias towards USA PIs could not be tested as data pre-dated 2025. Furthermore, changes in telescope allocation committee policy and the potential effect of telescope overhead changes were not fully assessed. The current research note provides an update by using the latest Cycles and expands the sample to Cycles 1 – 5 (n=1208 JWST programs) to provide tentative answers to these unresolved issues. While there is strong support to the overall findings for Cycles 1 – 3, we now note the emergence of USA PI and prior JWST experience as positive determinants of time allocation in regression models. The updated analysis also provides new insights concerning bias in the extra-galactic scientific categories.

## 1. Introduction

The allocation of time to astronomers to use telescopes for space research is critically important not only for new discoveries and the scientific field to progress, but also for individuals' careers and reputations of institutions that employ them. In the context of intense competition in science (Felt, 2017; Rao, 2024; Seeber et al., 2019), astronomers engage in competition to win time on shared telescope assets. This competition for time has been shown in previous research to be subject to different forms of bias (Langford and Langford, 2000; McCray, 2000; Reid, 2014). Williams (2025) examined potential sources of bias in relation to time allocation on the James Webb Space Telescope (JWST) over the first three Cycles of its operation, in 2022, 2023 and 2024, through the lens of moral economy (Atkinson-Grosjean and Fairley, 2009). Analysis of General Observer (GO) program data was supportive of a less biased institutional environment than may have been present in space research in the past, using allocated time as a measure of reward.

At the time of publication of Williams (2025), only the first three Cycles of JWST program data were publicly available, enough to reach conclusions about initial usage of the most expensive and in-demand space telescope ever built (Rao, 2024). Certain issues remained as recommendations for future research as and when subsequent Cycles of data would be made available. The main unresolved issues were as follows:

### *1.1 Unresolved issue #1: A return to long-standing norms of gender bias towards male PIs?*

The first issue related to the coefficient for PI gender in regression models across the first three Cycles. While this was insignificant in the full model, differences were found when the data was partitioned by Cycle. There was evidence of potential bias towards female PIs for the initial Cycle (Cycle 1), and back towards male PIs for Cycle 3. If this trend continued, it would have implications for how we understand time allocation on newly deployed mega-science platforms, and whether long-standing concerns relating to gender bias are to be upheld as these platforms become operational and more competitive over time.

### *1.2 Unresolved issue #2: Would there be any impact of the incoming Trump administration on asset usage decisions?*

The second issue related to the influence of administration change, given changes to Diversity, Equity, and Inclusion (DEI) policy, funding shake-ups in federal research, and national priority emphasis. It was unknown whether the broader political narrative could influence time usage on a telescope asset that was primarily US funded and operated by the Space Science Telescope Institute (STScI) in Baltimore. Scientific proposals for Cycles 4 and 5 were evaluated in telescope allocation committee (TAC) meetings within the context of the opening period of the second Trump administration.

### *1.3 Unresolved issue #3: Are findings based on new Cycles robust to changes in internal asset allocation policy?*

A third issue that was not fully possible to evaluate using the first three Cycles of GO data was whether changes in policy for evaluating scientific merit of proposals for JWST usage would have any effect on outcomes. Changes could result in different patterns of coefficients when comparing only Cycle 1 – 3 models with only Cycle 1 – 5 models, or by comparing Cycles 4 and 5 coefficients with previous cycles.

In addition to these unresolved issues, a minor - more technical - question had arisen during the review process for Williams (2025). This concerned the status of accepted GO programs. Accepted programs can take different status categories, including: "Pending submission", "Implementation", "Flight Ready", "Scheduled" and "Completed". Allocated time is sometimes adjusted before program completion for overhead changes and other operational reasons. While "Completed" status was controlled for in regression models, an update to the status variable when more programs had reached "Completed" status would help to establish whether these operational adjustments could have a material effect on interpretation of regression results. In Williams (2025), n=502 programs had been completed. In the analysis below, n=687 had been completed.

## 2. Methodology

The same method that was used in Williams (2025) was used in the current analysis, with a mix of bi-variate correlations, regression models, tabulated output, and margins plots on programs that were not classified as archival research and with time allocated as the dependent variable, log transformed to eliminate heteroskedasticity. The following steps were taken. Firstly, Cycles 4 and 5 data was collected from STScI public sources and appended to the previous dataset, and all variables created for Cycles 4 and 5 as reported in Williams (2025), including PI gender, team size, team geographic diversity, and various dichotomous flags such as institution and country flags. A description of all variables is given in Appendix A. Secondly, it was noted that the grouping of scientific categories for proposals was updated by STScI between Cycles 3 and 4. As the analysis involved a control variable (*Distance cat*) that proxied for the distance of the target object from the Earth, a revised mapping was created to map the new scientific categories onto *Distance cat* for Cycle 4. This was a straightforward exercise as some categories remained the same (e.g., Solar System Astronomy was unchanged and was mapped onto *Distance cat* = 1), while other altered scientific category labels could be mapped easily (e.g., Stars and Stellar Populations was new in Cycle 4, replacing Stellar Populations and the Interstellar Medium and Stellar Physics and Stellar Types in Cycle 3, and this was mapped onto *Distance cat* = 2 [own galaxy]). Thirdly, the program status variable (*Completed flag*) was updated for all programs, including those in Cycles 1 – 3. This was done on 21$^{st}$ May 2026, while that for Williams (2025) was done on 1$^{st}$ January 2025. This had the effect of capturing more completed programs since Williams (2025) and allowed a robustness test on whether overhead changes would have any material impact on the regression models for time allocated.

### 3. Updated results

The full sample size for the revised analysis is n=1208 JWST programs, compared to n=723 for the Cycles 1 – 3 analysis. Table 1 shows the revised descriptive statistics and Table 2 shows the revised correlation matrix. The addition of Cycles 4 and 5 data has the effect of increasing mean *Team size* (15.44 to 15.99) and geographic diversity of teams (*Geo diversity*) (Shannon Index increasing from 0.94 to 0.97). JWST teams are getting larger and more diverse. It also indicates that the fraction of programs waiving exclusivity drops as Cycles 4 and 5 are added (mean of *Non-exclusive* is 0.13, compared to 0.19 previously). *JWST experience* increases from 0.22 to 0.33. PIs with prior successful experience on JWST are winning more time in future proposals. The correlation matrix for Cycle 1 - 5 (Table 2) is very comparable to that reported for Cycle 1 – 3. For instance, *Team size* and *Non-exclusive* continue to have the highest positive bi-variate associations with allocated time. Larger and more geographically diverse teams are positively associated with high values of *Distance cat*, i.e., extra-galactic programs examining objects deeper and further in space and time. These coefficients have very similar strengths to the previous analysis. US affiliated PIs continue to be associated with less geographically diverse teams (r=-0.45, compared to r=-0.48 previously).

| | Mean | Std | Min | Max |
|---:|---|---|---|---|
| Allocation 1 | 34.33 | 58.67 | 0.4 | 1000 |
| PI gender 2 | 0.67 | 0.47 | 0 | 1 |
| SGJ PI 3 | 0.09 | 0.29 | 0 | 1 |
| USA PI 4 | 0.59 | 0.49 | 0 | 1 |
| Team size 5 | 15.99 | 11.98 | 1 | 100 |
| Geo diversity 6 | 0.97 | 0.61 | 0 | 2.41 |
| Non-exclusive 7 | 0.13 | 0.33 | 0 | 1 |
| JWST experience 8 | 0.33 | 0.47 | 0 | 1 |
| Mode count 9 | 1.35 | 0.61 | 1 | 5 |
| Instr. Count 10 | 1.34 | 0.54 | 1 | 3 |
| Distance cat 11 | 2.34 | 0.60 | 1 | 3 |
| Cycle 12 | 2.96 | 1.44 | 1 | 5 |
| Completed 13 | 0.57 | 0.49 | 0 | 1 |
| TOO 14 | 0.03 | 0.18 | 0 | 1 |

**Table 1.** Descriptive statistics Cycles 1 - 5

| | 1 | 2 | 3 | 4 | 5 | 6 | 7 | 8 | 9 | 10 | 11 | 12 | 13 |
|---|---|---|---|---|---|---|---|---|---|---|---|---|---|
| Allocation (ln) 1 | | | | | | | | | | | | | |
| PI gender 2 | -0.00 | | | | | | | | | | | | |
| SGJ PI 3 | 0.03 | -0.01 | | | | | | | | | | | |
| USA PI 4 | 0.06 | -0.03 | 0.27 | | | | | | | | | | |
| Team size 5 | 0.35 | 0.03 | 0.03 | -0.09 | | | | | | | | | |
| Geo diversity 6 | 0.14 | 0.03 | -0.17 | -0.45 | 0.54 | | | | | | | | |
| Non-exclusive 7 | 0.33 | 0.01 | 0.07 | 0.07 | 0.37 | 0.07 | | | | | | | |
| JWST experience 8 | 0.14 | 0.04 | -0.00 | 0.11 | 0.05 | -0.02 | 0.06 | | | | | | |
| Mode count 9 | 0.04 | 0.02 | 0.02 | 0.07 | 0.07 | 0.02 | 0.08 | 0.01 | | | | | |
| Instr. count 10 | 0.05 | 0.01 | 0.05 | 0.05 | 0.10 | 0.02 | 0.10 | 0.03 | 0.85 | | | | |
| Distance cat 11 | 0.13 | 0.01 | -0.09 | -0.14 | 0.28 | 0.26 | 0.06 | -0.03 | -0.04 | -0.02 | | | |
| Cycle 12 | 0.09 | -0.01 | -0.00 | 0.01 | 0.04 | -0.02 | -0.02 | 0.33 | -0.04 | -0.01 | -0.01 | | |
| Completed 13 | -0.15 | -0.01 | -0.03 | -0.03 | -0.07 | 0.02 | -0.04 | -0.26 | -0.04 | -0.08 | 0.02 | -0.81 | |
| TOO 14 | -0.03 | 0.06 | 0.04 | 0.01 | 0.12 | 0.04 | 0.19 | 0.01 | 0.18 | 0.19 | -0.09 | -0.01 | -0.11 |
| | | | | | | | | | | | | | |

**Table 2.** Correlation matrix Cycles 1 - 5

Table 3 shows the revised main regression models. In the full model (left most column), we note a very similar pattern of coefficients to Williams (2025), except for the coefficient for *USA PI*. *Team size* and *Non-exclusive* are the two strongest predictors of time allocated within the main group of independent variables, after accounting for all other variables of interest. This pattern is very consistent as the data is partitioned into the five Cycles, although we notice a drop in significance for *Non-exclusive* for Cycle 4. There is an almost identical pattern of coefficients for the *Distance cat* models compared to the previous analysis also, apart from *USA PI* becoming significant for *Distance cat* = 3 (extra-galactic programs). Previously this was positive but not significant. The coefficient for *Non-exclusive* for the partition *Distance cat* = 1 (solar system astronomy) is now significant at the $p<0.01$ level. We see non-significant coefficients for *PI gender* in Cycles 4 and 5, suggesting that a swing back to the long-standing norm for gender bias as far as this platform is concerned has not happened (unresolved issue #1 above).

A new development in the main regression models concerns the strength of *JWST experience*. This takes on much greater importance now Cycles 4 and 5 data is added, with its significance in the full model moving from $p=0.11$ to $p<0.01$. It is also positive and significant for the Cycle 4 partition, as well as for *Distance cat* = 3 ($p<0.01$).

| Variable | Full model | Cycle = 1 | Cycle = 2 | Cycle = 3 | Cycle = 4 | Cycle = 5 | Distance cat = 1 | Distance cat = 2 | Distance cat = 3 |
|---|---|---|---|---|---|---|---|---|---|
| *Independent* | | | | | | | | | |
| PI gender | -0.02 (0.05) | -0.18+ (0.09) | -0.08 (0.11) | 0.21+ (0.12) | -0.06 (0.12) | 0.07 (0.13) | -0.07 (0.18) | -0.14* (0.07) | 0.19* (0.08) |
| SGJ PI | -0.05 (0.09) | -0.13 (0.16) | -0.02 (0.18) | -0.27 (0.19) | 0.06 (0.24) | 0.07 (0.22) | -0.32 (0.26) | 0.07 (0.12) | -0.13 (0.16) |
| USA PI | 0.13* (0.06) | 0.11 (0.11) | -0.01 (0.13) | 0.17 (0.15) | 0.03 (0.16) | 0.26+ (0.14) | 0.14 (0.32) | 0.10 (0.08) | 0.20* (0.09) |
| Team size | 0.02*** (0.003) | 0.03*** (0.004) | 0.02** (0.01) | 0.04*** (0.01) | 0.02** (0.01) | 0.01* (0.01) | 0.05* (0.02) | 0.02*** (0.004) | 0.02*** (0.003) |
| Geo diversity | 0.02 (0.06) | -0.03 (0.09) | -0.02 (0.14) | -0.07 (0.14) | -0.12 (0.14) | 0.20 (p=0.12) (0.13) | -0.06 (0.23) | -0.01 (0.07) | 0.07 (0.09) |
| Non-exclusive | 0.68*** (0.11) | 0.62*** (0.19) | 0.98*** (0.23) | 0.46* (0.22) | 0.45 (0.34) | 0.75** (0.27) | 1.02** (0.46) | 0.34+ (0.18) | 0.97*** (0.15) |
| *Control* | | | | | | | | | |
| JWST experience | 0.17** (0.05) | 0.00 (0.17) | 0.32** (0.12) | 0.08 (0.12) | 0.25* (0.12) | 0.12 (0.13) | -0.07 (0.21) | 0.12 (0.08) | 0.26** (0.09) |
| Mode count | 0.02 (0.08) | 0.29* (0.15) | -0.05 (0.14) | -0.57*** (0.17) | 0.22 (0.18) | 0.08 (0.19) | 0.27 (0.20) | -0.04 (0.10) | 0.08 (0.13) |
| Instr. Count | 0.00 (0.09) | -0.31 (0.21) | -0.10 (0.17) | 0.55** (0.10) | -0.14 (0.18) | 0.08 (0.20) | -0.32 (0.24) | 0.16 (0.11) | -0.20 (0.16) |
| Distance cat | 0.06 (0.04) | -0.00 (0.08) | 0.07 (0.11) | 0.05 (0.10) | -0.01 (0.10) | 0.18+ (0.10) | Omitted | Omitted | Omitted |
| Cycle | -0.06+ (0.03) | Omitted | Omitted | Omitted | Omitted | Omitted | -0.15 (0.09) | 0.02 (0.04) | -0.08 (0.05) |
| Completed | -0.35*** (0.09) | 0.00 (0.51) | -0.07* (0.25) | -0.58** (0.20) | -0.64** (0.24) | Omitted | -0.46* (0.22) | -0.27* (0.12) | -0.42** (0.16) |
| TOO | -0.69*** (0.14) | -0.78* (0.39) | -0.78* (0.33) | -0.96** (0.31) | -0.54 (p=0.11) (0.34) | -0.19 (0.26) | -0.65** (0.23) | -0.59*** (0.18) | -0.41 (0.42) |
| Cons. | 2.69*** | 2.59*** | 2.61*** | 2.52*** | 2.76** | 1.71*** | 3.13*** | 2.60*** | 2.96*** |
| F | 20.09 | 7.42 | 6.44 | 9.24 | 4.11 | 5.02 | 2.97 | 7.07 | 16.28 |
| R2 | 0.21 | 0.27 | 0.24 | 0.36 | 0.16 | 0.20 | 0.30 | 0.15 | 0.29 |
| N | 1208 | 265 | 242 | 216 | 243 | 242 | 81 | 635 | 492 |

+ p<0.10, * p<0.05, ** p<0.01, *** p<0.001; robust standard errors in parenthesis

**Table 3.** Regression analysis for Cycles 1 – 5: main models

Tables 4 and 5 show updated regression models for partitions based on time allocated and *PI gender*. The partitions based on time allocated (Table 4) are selected using the mean value in the new sample (34.33 hours). The *PI gender* coefficients in Table 4 have the same sign as with the Cycles 1 – 3 analysis, although significance is weaker. This can be instructive; while the full model in Table 3 indicates no gender bias, different decision processes are used for different program size tranches. Whether these different decision processes are linked to the contrasting *PI gender* coefficients should be investigated in future work. While proposals are strictly anonymized on submission, reducing conscious bias, the persistence of this pattern is a possible indication of unconscious bias within an institutional framework that has differentiated evaluation processes. Interestingly, *Team size* is no longer significant for the partition of larger programs. *Non-exclusive* is not significant for the partition of smaller programs, in keeping with the previous analysis. *USA PI* becomes significant for smaller programs; not the case in the previous analysis. *JWST experience* becomes significant in smaller and larger program partitions; previously only smaller programs.

| Variable | All 5 Cycles | |
|---|---|---|
| *Independent* | Allocation < = 34.33 hours | Allocation > 34.33 hours |
| PI gender | -0.06<br>(0.04) | 0.10+<br>(0.06) |
| SGJ PI | 0.01<br>(0.08) | -0.04<br>(0.08) |
| USA PI | 0.10*<br>(0.05) | 0.01<br>(0.07) |
| Team size | 0.02***<br>(0.003) | 0.003<br>(0.003) |
| Geo diversity | -0.01<br>(0.05) | -0.05<br>(0.07) |
| Non-exclusive | -0.13<br>(0.10) | 0.70***<br>(0.09) |
| *Control* | | |
| JWST experience | 0.11*<br>(0.05) | 0.15*<br>(0.06) |
| Mode count | 0.07<br>(0.06) | -0.004<br>(0.06) |
| Instr. Count | 0.01<br>(0.08) | -0.04<br>(0.07) |
| Distance cat | -0.03<br>(0.04) | 0.08 (p=0.11)<br>(0.05) |
| Cycle | -0.06*<br>(0.03) | -0.02<br>(0.04) |
| Completed | -0.07<br>(0.08) | -0.11<br>(0.11) |
| TOO | -0.19<br>(0.13) | -0.40*<br>(0.20) |
| Cons. | 2.51*** | 3.86** |
| F | 5.01 | 10.46 |
| R2 | 0.06 | 0.34 |
| N | 892 | 316 |

+ p<0.10, * p<0.05, ** p<0.01, *** p<0.001; robust standard errors in parenthesis

**Table 4.** Regression analysis Cycles 1 – 5: program size partitioned

Table 5 shows that the effects of the main variables of interest across the two gender partitions are the same in the updated analysis with all five Cycles, compared to the previous analysis with three Cycles. Regardless of whether the PI was male or female, the main predictors of time allocated are still *Team size* and *Non-exclusive*. However, we do now note the positive and significant coefficient for *USA PI* for the female partition ($p<0.05$), but not the male partition. This was not the case with the Cycles 1 – 3 analysis. All other control variables are similar in terms of their effects, except for *JWST experience*. As with the main regression models in Table 3, we now see a stronger effect for *JWST experience* across both time allocated partitions (Table 4), and the PI gender = Male partition (Table 5). *JWST experience* is emerging as one of the most important determinants of allocated time in regression models with this effect being particularly strong for male PIs (Table 5) and extra-galactic programs (Table 3).

| Variable | PI gender | |
|---|---|---|
| *Independent* | Female | Male |
| PI gender | Omitted | Omitted |
| SGJ PI | 0.05<br>(0.13) | -0.06<br>(0.12) |
| USA PI | 0.24**<br>(0.11) | 0.08<br>(0.07) |
| Team size | 0.02***<br>(0.01) | 0.02***<br>(0.003) |
| Geo diversity | 0.15<br>(0.11) | -0.03<br>(0.07) |
| Non-exclusive | 0.61**<br>(0.20) | 0.72***<br>(0.14) |
| *Control* | | |
| JWST experience | 0.15<br>(0.10) | 0.19**<br>(0.07) |
| Mode count | -0.03<br>(0.16) | 0.03<br>(0.09) |
| Instr. Count | 0.01+<br>(0.17) | -0.00<br>(0.10) |
| Distance cat | -0.13+<br>(0.07) | 0.15**<br>(0.05) |
| Cycle | -0.07*<br>(0.05) | -0.05<br>(0.04) |
| Completed | -0.27+<br>(0.16) | -0.38***<br>(0.11) |
| TOO | -0.51<br>(0.33) | -0.71***<br>(0.15) |
| Cons. | 2.97*** | 2.50*** |
| F | 6.86 | 17.09 |
| R2 | 0.21 | 0.223 |
| N | 400 | 808 |

+ $p<0.10$, * $p<0.05$, ** $p<0.01$, *** $p<0.001$; robust standard errors in parenthesis

**Table 5.** Regression analysis Cycles 1 – 5: PI gender partitioned

Figure 1 shows the margins plots for: *PI gender* across Cycles and distance categories. Figure 2 shows the margins plots for *USA PI* across Cycles and distance categories. The convergence of the dotted blue line for female PIs with the solid red line for male PIs in Figure 1 reinforces the results in Table 3 in terms of the *PI gender* coefficient across the Cycles. There is no evidence of a swing back to a long-standing norm for gender bias towards male PIs in the overall dataset. However, there is some evidence of an evolution of *PI gender* bias toward male PIs in the extra-galactic programs (*Distance cat* = 3). The significance level for this is stronger in the current analysis (p<0.05), compared to the Cycles 1 – 3 analysis (p<0.1). The stronger effect of *USA PI* is reflected in Figure 2 as the Cycles proceed and is also seen to be more prominent for extra-galactic programs.

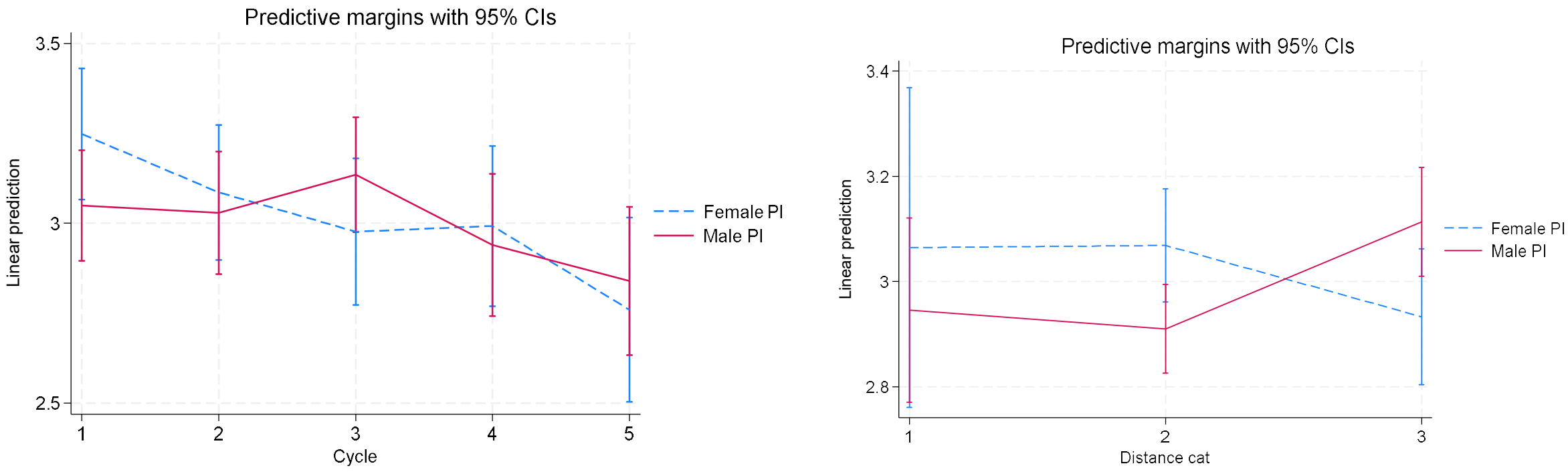

**Figure 1.** Interaction between *PI gender* and (1) *Cycle* (left), and (2) *Distance category* (right) (n=1208)

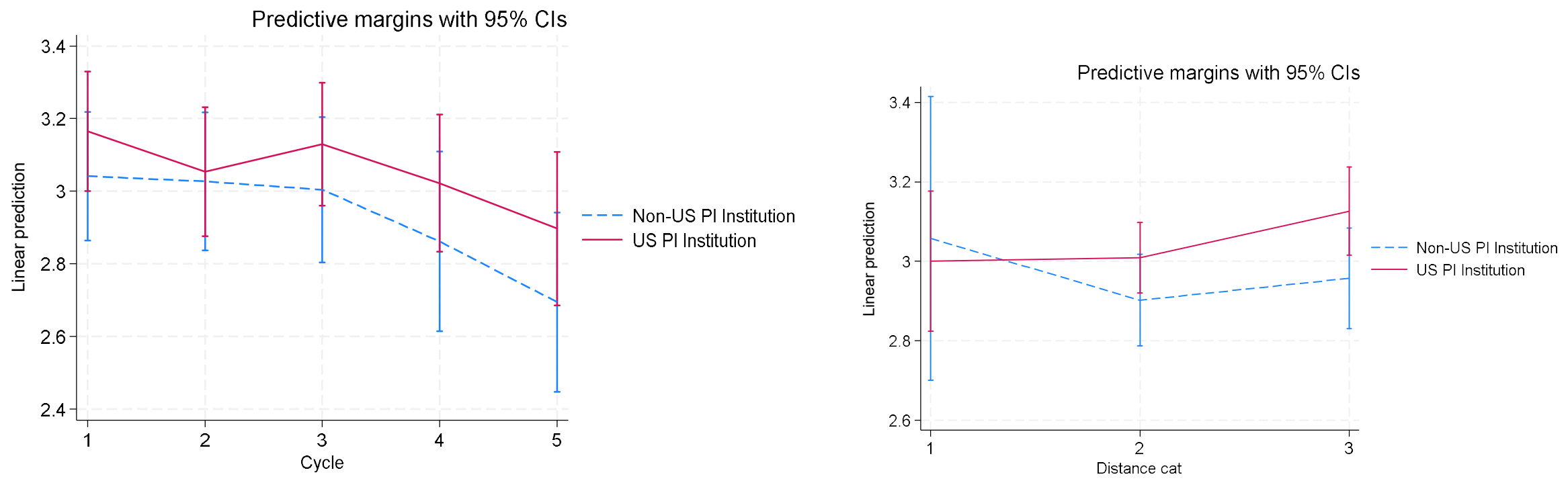

**Figure 2.** Interaction between *USA PI* and (1) *Cycle* (left), and (2) *Distance category* (right) (n=1208)

To summarize, in terms of the three unresolved issues listed above, we can now provide answers as follows.

### *3.1 Unresolved issue #1: A return to long-standing norms of gender bias towards male PIs?*

Within the whole dataset, the question of a swing-back to male PIs having a positive association with time allocated as Cycles progress receives no support (Table 3, left-most model and Figure 1, left). What is interesting concerning *PI gender*, however, is continued support for the previous analysis in terms of its effect in partitions of the data by time allocation categories, broadly defined as: "smaller" vs. "larger." Opposite signs for the coefficients for *PI gender* are still seen here when Cycles 4 and 5 are added. The situation is similar with the interaction with *Distance cat* (Table 5 and Figure 1, right) where there appears to be a growing bias towards male PIs for extra-galactic programs (*Distance cat* = 3). There is no *overall* return to a long-standing norm of gender bias against female PIs, but there is a persistent difference when program categories are analysed separately in these partitions.

### *3.2 Unresolved issue #2: Would there be any impact of the incoming Trump administration on asset usage decisions?*

The most consistent predictors are still *Team size* and *Non-exclusive*, while the flag for *SGJ PI* (whether the PI was from STScI, the NASA Goddard Space Flight Center or Johns Hopkins University) remains non-significant. However, we do note that *USA PI* (whether the PI was based in a US institution as listed on the program log) now turns positive and significant in the full model when Cycles 4 and 5 are included and has a significant coefficient in Cycle 5 (Table 3, and Figure 2, left). Descriptive analysis reveals an increase in the fraction of USA PI led programs in Cycles 4 and 5 compared to the previous three Cycles. Expressed as a percentage, the level for USA PI led programs in Cycles 4 and 5 was 60.2% while in Cycles 1 – 3 it was 57.6%. This increase in the fraction of USA PI led programs is small and not a sign of a sudden strengthening of national prioritization. However, the new coefficient for USA PI in regression models, and the pattern in Figure 2, could be suggestive of such a trend and is worth following up in future Cycles.

### *3.3 Unresolved issue #3: Are findings based on new Cycles robust to changes in internal asset allocation policy?*

Changes in time allocation policy between Cycles 3 and 4 included the different labelling for scientific categories as noted above. There were still eight categories, and these were mapped in the current analysis to the same *Distance cat* variable. There was also a change to the definition of proposal category by size. In Cycle 3, the categories were smallest (<15 hours), small (15-25 hours), medium (>25 to 75 hours), and large (> 75 hours). In Cycle 4, this changed to smallest (< 20 hours), small (20 to <=50 hours), medium (>50 to <=130 hours), and large (>130 hours). The smallest category proposals were graded independently by external reviewers, not in panels, in Cycles 3 and 4. The small and medium categories were assessed by panels meeting online, and the large category was assessed by the executive committee meeting in person. The main change here was in the delineation of the category boundaries. An emerging theme is the increased prominence of *JWST experience* in regression models. This is now significant in the full model, as well as in the partition for *Distance cat* = 3 (extra galactic). This is interesting because success in previous time allocation competitions on the same platform could inspire knowledge, insight and confidence to push forward with larger and more ambitious proposals with a greater appetite for risk. This experience effect appears to align with the shifting program size boundaries as the Cycles progressed and while competition for time became fiercer. JWST proposal rules do not allow proposers to identify themselves; they would be rejected. But they are allowed to refer to previous approved programs, even citing program IDs. It is plausible that members of the STScI community can reliably identify submitting PIs based on content of proposals when previous programs are cited, whilst also having knowledge of the success of any previous programs. The strengthening of the JWST experience variable could be suggestive that ongoing rule changes regarding reviews and program sizes favour PIs with prior success on JWST.

The minor question was on whether the updating of the *Completed flag* status would have any material effect on the model. On re-running the model for Cycles 1 – 3 (n=723) after updating this flag on 21st May 2026, the full model pattern of coefficients is the same as that reported in Williams (2025) where the completed status was updated on 1st January 2025. Indeed, the model R2 moves from 0.24 to 0.25, and the pattern for all other coefficients, signs and significances stay the same. This implies that telescope overhead changes for operational reasons following proposal acceptance do not alter the interpretation of results. Future research on allocated time can confidently use accepted program data for programs that are scheduled but not yet completed.

## 4. Discussion

This research note provides an empirical update to the tests conducted in Williams (2025), an exploration of potential bias within the competition for time utilization of the James Webb Space Telescope (JWST) by the professional astronomy community. The current update includes the latest JWST GO program data, i.e., Cycles 4 and 5, research programs that started in July 2025, and an update to programs that have since been completed. Within the first three Cycles of usage, JWST made groundbreaking discoveries and new contributions to knowledge of the cosmic dawn (Napolitano et al., 2025), early forming black holes (Jeon et al., 2025), and exoplanets (Carter et al., 2024), as well as in many other areas. The fact that greater openness and less bias had been detected (compared to what may have happened in the past) and that this is linked to these outcomes, demonstrates the value of norms for openness and removing bias in the institutional framework for time allocation.

Gender bias, as captured through the gender of the PI having a significant beta coefficient in regression modelling of allocated time, was not found to swing towards males when Cycles 4 and 5 were added, refuting a suggestion that arose in the Cycles 1 – 3 analysis. This suggests gender equality interventions and awareness have been sincere (Meyer et al., 2025; Primas, 2021) and have permeated the institutional environment for time allocation on this platform. Findings do, however, continue to show different effects for *PI gender* when time allocation data is partitioned (by (1) “smaller” vs. “larger” time allocation and (2) “solar system astronomy” vs. “galactic astronomy” vs. “extra-galactic astronomy”). Future work is needed here to isolate the reasons for this. It may be that interventions are easier to implement when less is at stake (“smaller” programs). It could also be that ‘super-star’ astronomers with the convening power to bring together large investigating teams able to successfully bid for larger time allocations for the most challenging objects historically tended to be male, and that it will take time to re-balance following the current era of interventions.

The impacts of the Trump era policies, and the incremental change in platform policy for time allocation within STScI are examined by comparing Cycles 4 and 5 regression coefficients with previous Cycles, and by comparing those at the level of the expanded dataset with the earlier dataset. The lack of any major change in most bias predictors suggests the political zeitgeist, especially around DEI, nationalism, cultural identity and national priorities, has not affected the time allocation process for JWST. However, the strengthening of the coefficient for *USA PI* following the inclusion of Cycles 4 and 5 is worth tracking in future Cycles to understand how this is linked to rewards allocated. Future work can examine whether the percentage of US PI-led programs on JWST continues to increase. It can also look at the make-up of investigating teams, and whether these become more homogeneous, or indicative of national bias in the new era.

An additional new finding from the current analysis is the significance of *JWST experience*, i.e., whether the PI already had experience on a successful JWST program, as a determinant of future allocated time. This emerging finding can contribute to debates on PI level determinants of productivity in science. It is in line with prior studies on researcher's position and standing as determinants of grant outcomes (e.g., Grimpe, 2012), and on how project specific experience can be useful for future success through learning of the reward 'system', including persuasive proposal writing (McAlpine, 2020). Future work could delve into the relationship between experience and time allocation within the competitive environment of astronomy research in more granular detail, perhaps examining experiences within the investigating teams, as well as the mechanism by which successful experiences lead to increased competitiveness and proposals with higher levels of scientific merit.

## APPENDIX 1 – List of variables

| Variable name | Description | Source |
|---|---|---|
| Allocation | Allocation (in hours) per program, log transformed | Allocation stated in program information by proposal:<br>https://www.stsci.edu/jwst/science-execution/program-information?id=XXXX (where XXXX = Program ID)<br>21st May 2026 |
| PI gender | Dichotomous variable for gender of principal investigator (PI), 1 = Male, 0 = Female | Estimated through PI name from abstract catalogue cross referenced to institutional website and online references |
| SGJ PI | Institution of PI, 1 = Space Telescope Science Institute or NASA Goddard Space Flight Center or Johns Hopkins University, 0 otherwise | PI institution https://www.stsci.edu/jwst/science-execution/approved-programs/general-observers/cycle-1-go |
| USA PI | Country of PI institution listed on program abstract, 1 = USA, 0 otherwise | PI institution https://www.stsci.edu/jwst/science-execution/approved-programs/general-observers/cycle-1-go |
| Team size | Complete count of investigators per program | List of investigators in program information by proposal:<br>https://www.stsci.edu/jwst/science-execution/program-information?id=XXXX (where XXXX = Program ID) |
| Geo diversity | Shannon entropy measure of diversity of country of institutions of all investigators for each program<br>H = -Σpi * ln(pi)<br>where pi is the proportion made up of group (country) i for the program | Country of institution of all investigators in program information by proposal:<br>https://www.stsci.edu/jwst/science-execution/program-information?id=XXXX (where XXXX = Program ID) |
| Non-exclusive | Dichotomous variable derived from exclusive access period (days) on Cycle 1-5 program information takes a value of 1 if exclusive access period = 0 (as opposed to 3, 6 or 12 months) | Exclusive access period<br>Cycle 1 :<br>https://www.stsci.edu/jwst/science-execution/approved-programs/general-observers/cycle-1-go<br>Cycle 2 :<br>https://www.stsci.edu/jwst/science-execution/approved-programs/general-observers/cycle-2-go |

| Variable name | Description | Source |
|---|---|---|
| | | Cycle 3 :<br>https://www.stsci.edu/jwst/science-execution/approved-programs/general-observers/cycle-3-go<br>Cycle 4 :<br>https://www.stsci.edu/jwst/science-execution/approved-programs/general-observers/cycle-4-go<br>Cycle 5 :<br>https://www.stsci.edu/jwst/science-execution/approved-programs/general-observers/cycle-5-go |
| JWST experience | Dichotomous variable indicating if the PI already led a previous GO program on JWST (=1, otherwise 0) | Derived from GO program information ordered chronologically by PI |
| Mode count | Count of distinct modes listed on program information (MRS, IFU, Coronagraphy, BOTS, Imaging, SOSS, LRS, AMI, FS) | Modes per program (sources for Cycles 1 – 5 same as previous row above) |
| Instrument count | Count of distinct instruments used (NIRSpec, NIRCam, MIRI, Wfsc Fine Phasing) | Instruments per program (sources for Cycles 1 – 5 same as previous row above) |
| Scientific category | Categorical variable ranging from 1 – 8 for each of the program groups | Sources for Cycles 1 – 5 same as previous row above |
| Distance category | Categorical variable mapping scientific category onto ordinal variable 1 = Own solar system, 2 = Own galaxy (including local group), 3 = Universe beyond own galaxy | Derived from scientific category |
| Cycle | Id for Cycle (1 – 5) | https://www.stsci.edu/jwst/science-execution/approved-programs |
| Completed | Dichotomous flag indicating if program was completed as of 21st May 2026 | https://www.stsci.edu/jwst/science-execution/program-information?id=XXXX (where XXXX = Program ID) |
| TOO | Dichotomous variable indicating if proposal is described as a Target of Opportunity (TOO) in the proposal type (=1, 0 otherwise) | https://www.stsci.edu/jwst/science-execution/program-information?id=XXXX (where XXXX = Program ID) |